\documentclass[%
 reprint,%
 amssymb, amsmath,%
 aip,cha, numerical%
]{revtex4-1}

\usepackage{bm}%
\usepackage[colorlinks=true,linkcolor=blue]{hyperref}%
\usepackage{graphicx}
\expandafter\ifx\csname package@font\endcsname\relax\else
 \expandafter\expandafter
 \expandafter\usepackage
 \expandafter\expandafter
 \expandafter{\csname package@font\endcsname}%
\fi
\usepackage{subcaption}
\usepackage{setspace}
\usepackage{color}

\begin{document}

\title{Short time dynamics determine glass forming ability in a glass transition two-level model: a stochastic approach using Kramers' escape formula}%

\author{J. Quetzalc\'oatl Toledo-Mar\'in}%
\affiliation{Departamento de Sistemas Complejos, Instituto de
F\'{i}sica, Universidad Nacional Aut\'{o}noma de M\'{e}xico (UNAM),
Apartado Postal 20-364, 01000 M\'{e}xico, Distrito Federal,
M\'{e}xico}%

\author{Gerardo G. Naumis}%
\email{naumis@fisica.unam.mx}
\affiliation{Departamento de Sistemas Complejos, Instituto de
F\'{i}sica, Universidad Nacional Aut\'{o}noma de M\'{e}xico (UNAM),
Apartado Postal 20-364, 01000 M\'{e}xico, Distrito Federal,
M\'{e}xico}%

\date{December 2016}%
\revised{?}%

\maketitle

\begin{quotation}
  The relationship between short and long time relaxation dynamics  is obtained for a simple solvable two-level energy landscape model of a glass. This is done through means of the Kramers transition theory, which arises in very natural manner to calculate transition rates between wells. Then the corresponding stochastic master equation is analytically solved to find the population of metastable states. A relation between the cooling rate, the characteristic relaxation time and the population of metastable states is found from the solution of such equation. From this, a relationship between the relaxation times and the frequency of oscillation at the metastable states, i.e., the short time dynamics is obtained. Since the model is able to capture either a glass transition or a crystallization depending on the cooling rate, this gives a conceptual framework in which to discuss some aspects of rigidity theory.
 \end{quotation}


\section{Introduction}
Despite the great use of glass in our societies; e.g. window glasses, smart-phone glasses, memory devices, optic fiber, containers, to name a few; glass transition has been proven to be  a very complex problem. Although a lot of progress has been made in the last half century, still there are many unsolved questions  \cite{dyre2006col, ngai, trachenko2011heat, Dyre1987, Dyre1995, langer1988entropy, naumis2006variation, phillips1996stretched, naumis1998stochastic, micoulaut1999glass, kerner2000stochastic, Mauro1, Mauro2, mauro3PNAS, debenedetti1996metastable, debenedetti2001supercooled, stillinger2002energy, wang2001sharp, phillips1979topology, naumis2005energy, sastry, mezard2012glasses, gleim1998does, gleim2000relaxation, goldstein1969viscous, adam1965temperature, debenedetti2001theory, milchev1983influence, avramov1988effect, stillinger1999exponential, widmer2006predicting, faraone2004fragile, trachenko2015collective, dyre2016simple, doliwa2003energy, broderix2000energy, macedo1966two, matyushov2005two, angell1972configurational, tanaka1999two, debenedetti2003model, huse1986residual, langer1990nonequilibrium, langer1988entropy, brey1991residual}. What is even more interesting is that some of these questions are also present in other phenomena, for instance, protein folding,
turbulence and cell motion inside dense tissues \cite{dauchot2014glass, wales2000energy, manning, daggett2003there}. Hence, the growing appeal around glassy systems.

It is quite fair to say, from a technological and fundamental standpoint, that the most important variable for glass formation is the cooling speed \cite{Mauro1, Mauro2}. In his iconic paper \cite{phillips1979topology}, Phillips presents a dependence between the chemical composition and the minimal cooling speed necessary for glass formation for several chalcogenide alloys through means of his rigidity theory, which was later generalized by Thorpe \cite{thorpe1983continuous}. One of the main features in this theory may be summarized in the following manner: When the number of bond constraints equals the number of degrees of freedom, the glass forming ability is optimized, i.e., producing glass requires the slowest cooling rate. In this situation, the mean coordination number equals the critical percolation coordination number, i.e., domains of floppy modes (zero frequency modes) and rigid modes coexist. As the mean coordination number decreases, which may be tuned by varying the chemical composition, floppy mode domains grow while rigid mode domains disappear. As floppy modes increase in number, the glass formation is more difficult. In this sense, it has been well established theoretically and experimentally that isostatic rigid glasses are easier to form \cite{selvanathan2000stiffness, wang2001sharp, naumis2005energy, phillips1979topology, huerta}. Despite this, the glass formation dependence on the cooling rate is still poorly understood. Among the vast set of tools used to study supercooled liquids and glass transition is the energy landscape picture \cite{goldstein1969viscous}, however, it is not trivial to understand how the energy landscape depends upon the interatomic or intermolecular potential, and thus how the cooling rate is related with the topological sampling.

Another intriguing problem in super-cooled liquids is the relation between short  and long time dynamics \cite{widmer2006predicting, ngai, dyre2006col}. It is a known fact that a super-cooled liquid increases its viscosity or, equivalently, its relaxation time by more than ten orders of magnitude when the temperature is varied by a factor of three and, depending on this behavior, the supercooled liquid is called strong or fragile. Moreover, there are cases, for instance confined supercooled water \cite{faraone2004fragile}, where there is a transition from fragile-to-strong. But a connection between this feature and microscopic time dynamics is missing precisely because it is very difficult to establish a connection between processes on the picosecond time scale and on the second or larger time scale, least to say a causality. This is in part the reason why the question \textit{ to what extent are the long time dynamics determined by the short time dynamics?} is still an unanswered one. However, a flow event or molecular rearrangement in real space occurs on a very short time scale. These events correspond to a barrier transition in the energy landscape picture. This idea has been the starting point in the path to solve the aforementioned question, yet more work is needed \cite{dyre2006col}. 

\begin{figure}[h]
\centering
\includegraphics[width=2.5in]{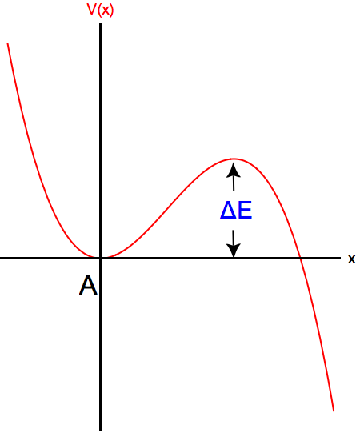}
\caption{\footnotesize Sketch of a potential $V(x)$ as a function of the reaction coordinate $x$, where $V(x)$ has a stable stationary point (A) and a barrier of height $\Delta E$.}\label{fig:kwell}
\end{figure}

From a different perspective, Kramers' transition state theory \cite{landauer1961frequency, zwanzig2001nonequilibrium, hanggi, coffey2004applications} gives a solid framework which can be used to study barrier transitions, at least as first means. This approach has been widely used in many different fields to understand how a system leaves an energy landscape basin \cite{hanggi,coffey2004applications}. These ideas put on a solid ground the empirical Arrhenius law, namely,

\begin{equation}
\tau (T)=\tau_0 \exp(\Delta E/kT) \; ,
\label{eq:Arrhenhius1}
\end{equation}

which relates the relaxation time   for leaving a basin  ($\tau(T)$) with the temperature ($T$).Here $\tau_0$ is the smallest oscillation period. As seen in figure \ref{fig:kwell}, $\Delta E$ is the energy barrier hill that closes the basin .  The escape over the barrier represents the breaking of a chemical bond  \cite{coffey2004applications}.

Kramers' formula indicates that the relaxation times, i.e., the escape time for leaving the basin is given by \cite{hanggi,coffey2004applications}, 

\begin{equation}
\tau(T)=\Lambda \frac{2 \pi}{\omega_{A}}  \exp(\Delta E/kT) \; ,\\
\label{eq:Kramers}
\end{equation}
where $\omega_{A}$  is the frequency of oscillation at the bottom of basin $A$ (see figure \ref{fig:kwell}) and $\Lambda$ is a factor that comes from a microscopic model of dissipation. The main improvement of Kramer's formula over the Arrhenhius one is the coupling with the thermal bath and the inclusion  of the energy basin oscillation period (which is related with the energy basin curvature) \cite{hanggi}.  Yet, on doing so one must consider the damping factor which appears in the prefactor of Kramers' transition rate given by Eq. (\ref{eq:Kramers}), which does not appears in the pure exponential Arrhenius formula Eq. (\ref{eq:Arrhenhius1}). 
This prefactor depends upon several considerations, like the dissipation regime (overdamped or underdamped).
Nonetheless, up to our knowledge, the effects of this prefactor  and its relationship with the basin curvature on glass relaxation  has not been considered . Some clear indications of such relationship has been found in molecular dynamics simulations \cite{flores2012mean} and theoretical analysis  \cite{dyre2006col, flores, flores2011boson}. Clearly, more work is needed due to its relevance \cite{mccann1999thermally, simon1992escape}.

The exploration of this subject is further complicated by the fact that there are many available models of glass transition \cite{Dyre1995}. Among these, two-level models are popular since they explain several essential thermodynamical properties of real glasses  \cite{macedo1966two,angell1972configurational, matyushov2005two}. From a more theoretical point of view, their appeal has never decreased since they serve as a starting point to develop simple stochastic models of glass formation  \cite{langer1989entropy, huse1986residual, langer1990nonequilibrium, langer1988entropy, brey1991residual}. Although, in general, these models lack for the complexity of the landscape, the problem can be solved by using a minimal model that incorporates a simple landscape topology \cite{naumis}. Furthermore, keeping in mind that cooling rate effects on glass formation is a poorly understood subject, one would expect that in any sensible glass transition model the phase transition to the crystal should be included for low cooling rates. In this sense, the aim in previous work was to present a landscape model with a minimal set of ingredients that would take this fact into account (see \cite{toledo2016minimal, naumis}) . The simple two-level system features a first order phase transition in the thermodynamical limit and, for some fast enough cooling rate, is able to arrest the system in metastable states mimicking the glass transition phenomena\cite{naumis}. The model relates the minimal cooling rate for a glass forming tendency with the thermal history, the energy landscape barrier and the characteristic relaxation time \cite{toledo2016minimal}. This last having an Arrhenius behavior. Moreover, this kind of two-level model can be put in correspondence with effective-mean field theories of glasses \cite{wolynes2012structural}. 
In our previous work we were not able to tackle the problem of how short and long time dynamics are related. In particular, we were interested on how the mean quadratic displacement, which is related with the curvature of the energy basin \cite{flores,flores2012mean}, determines glass relaxation. 

Here we explore how Kramers' transition state theory appears naturally in the model.  This allows us to relate the short and long time dynamics, while also allowing us to discuss some rigidity theory ideas. It is important to remark that our findings are present in all energy minima of the landscape, yet, a precise modelling for a real glass needs to consider other factors as we will see in the discussion section. 
The paper is organized as follows: In the following section we briefly present our model and its features in equilibrium. In section \ref{sec:quench} we study our model under a quench and discuss the glass formation tendency dependency with the short time dynamics. In section \ref{sec:char} we derive the characteristic relaxation time and discuss its dependence with the short time dynamics. In section \ref{sec:therm} we determine the minimal cooling rate for a strong glass forming tendency and discuss its dependence with short time dynamics. In section \ref{sec:Disc} we present all findings in our model in the context of rigidity theory, and from this we discuss the relation between transition barriers height and normal modes.  Finally, section \ref{sec:conclusions} presents the conclusions.

\section{Revisiting the glass transition two-level model}\label{sec2} 
\begin{figure}[!h]
\centering
\includegraphics[width=3.37in]{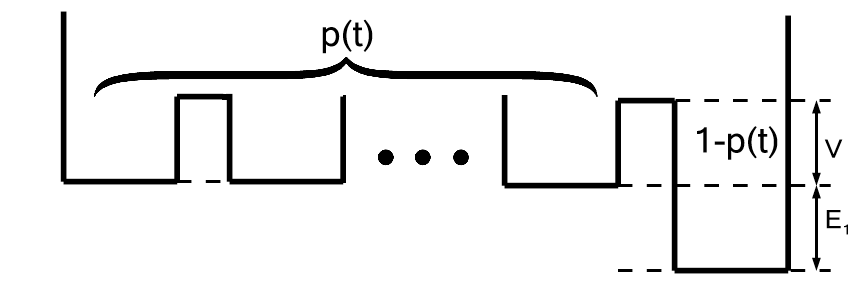}
\caption{\footnotesize The two level system energy landscape, showing the
barrier height $V$ and the asymmetry $E_1$ between the two levels. There are
$g_1$ wells with energy $E_1$, associated with metastable states, and $g_0$ ground states
with energy $E_0=0$. The population of the upper well is $p(t)$ \cite{naumis}.}\label{system}
\end{figure}

Let us, in a brief manner, define the original model to be used for the glass transition (for a detailed description see \cite{naumis}). As seen in Fig. \ref{system},  the energy landscape is composed by $g_1$ wells with energy $E_1=N\epsilon>0$ which we denote as the metastable states for $N$ atoms, and $g_0(\ll g_1)$ wells with energy $E_0=0$ which we denote as  ground states. All wells are interconnected, and any two wells have an in-between wall of height $V$.

\begin{figure}[h]
\centering
\includegraphics[width=3.35in]{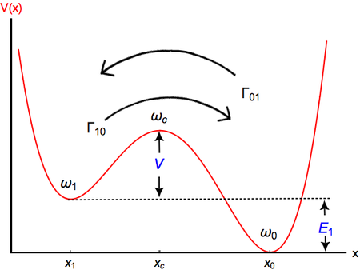}
\caption{\footnotesize Sketch of a double well potential with a barrier of height $V$, showing the frequencies associated with
each well and transition barriers. The transition rates between wells are also shown with arrows.}\label{fig:kwell1}
\end{figure}

Now, given this simple topology, we assume that transition probabilities between the the metastable states are all the same, in like manner transition probabilities between the ground states are all the same. The transition probability from each well in the metastable states to any of the ground states are all the same as well as the transition probability from each ground state well to any of the metastable states in the metastable state. The probability $p(t)$ of finding the system with high energy satisfies the following master equation\cite{naumis}:
\begin{equation}\label{master equation}
\dot{p}(t)=-\Gamma_{10}g_0 p\left(t\right)+\Gamma_{01}g_{1}\left(1-p\left(t\right)\right) \; ,
\end{equation}
where $\Gamma_{10}$ corresponds to the transition probability per time for going from a state with energy $E_1$ to a ground state, i.e., state with energy $0$, and  $\Gamma_{01}$ for the reverse transition. In the original model, both $\Gamma_{01}$ and  $\Gamma_{10}$ were assumed to be proportional to a common generic $\Gamma$, which was  the inverse frequency of oscillation on the wells, related with the curvature of the energy basin. It provided the time scale of the model. However, this was an oversimplification  since not all wells have the same oscillation frequency, which is a well known difference between glasses and crystals \cite{flores2012mean}. Thus, here we propose to use Kramers theory to take into account in a proper way such contribution. To do this, we consider that the square well model must be replaced by a smooth potential. Since the square well model can be reduced to a model of two levels with degeneracy, then the model with the smooth potential can be translated into a landscape with the shape shown in Fig. \ref{fig:kwell1} with the same degeneration as in the original square model.

According to Kramers' first passage time formulation in the overdamped scenario \cite{zwanzig2001nonequilibrium, hanggi, coffey2004applications}, one finds that 
\begin{equation}
\begin{cases}
\Gamma_{10}(T)=\frac{\omega_1 \omega_c }{2\pi \gamma} e^{-V/T} \; , \\
\Gamma_{01}(T)=\frac{\omega_0 \omega_c }{2\pi \gamma} e^{-(V+E_1)/T} \; , 
\end{cases}
\label{Gamas}
\end{equation}
where $\omega_1^2\equiv V''(x_1)/M$ is the squared angular frequency inside the metastable minimum at position $x=x_1$ and $M$ is the mass of the system. $V''(x)$ denotes the second derivative of the potential at $x$.
$\omega_0^2\equiv V''(x_0)/M$  is the squared angular frequency inside the global minimum at position $x=x_0$ and $\omega_c^2\equiv |V''(x_c)|/M$ is the squared angular frequency at the transitional state at position $x=x_c$ (see Fig. \ref{fig:kwell1}). Also, the energy potential barrier height that appears in Eq. (\ref{Gamas}) is given by $V$, which is defined as $V=V(x_c)-E_1$.

For real glasses, $\omega_1$ and $\omega_0$ depend upon the normal modes frequencies at the energy minima \cite{flores2012mean, flores2011boson, flores}. Here is important 
to remark that in general, the potential barrier height $V$ can be correlated  with the surrounding energy minima. 
In this sense, the normal modes frequencies and the energy minima of the energy landscape is known for a variety of 
interaction potentials \cite{sastry, Wales2}. Also, the energy barriers distribution \cite{doliwa2003energy} and the Hessian index \cite{broderix2000energy} as function of temperature has been obtained for Lennard-Jones supercooled liquids. Moreover, in a quite interesting report, Wales \cite{wales2001microscopic} proposes the use of catastrophe theory for the characterization of the energy landscape. Nonetheless, the relation between the transition barrier heights and the frequencies is still an open issue, but certainly they can be correlated \cite{hansen2016connection}. Throughout this paper,  $V$ is treated as an independent parameter that can be or not correlated with the frequency $\omega$ at the bottom of the adjacent wells. Here we only need a maximal $V(x)$, identified with $V$, and a finite $V''(x)$ at such point.  In that sense, our model describes a general situation of an energy barrier and two adjacent minima. Later on, in the discussion section,  we will  consider how to model real glasses, since one needs to include many energy minima and barriers between them. This needs to include correlations between energy minima, barrier height and curvature, as well as the fact that in real glasses, there is a {\it distribution of energy barriers, energy minima and curvatures} that are needed to be overimposed on our minimal picture of the landscape.

It is worthwhile mentioning that the transition rates shown in Eq. (\ref{Gamas}) satisfy the detailed balance condition,
\begin{equation}\label{15}
\frac{\Gamma_{01}}{\Gamma_{10}}=\frac{\omega_0}{\omega_1} e^{-E_{1}/T} \; ,
\end{equation}
Also, it is important to observe that the stationary solution to our master Eq. (\ref{master equation}) is:
\begin{equation}
p_0(T)=\frac{\frac{g_1 \omega_0}{g_0 \omega_1}e^{-E_1/T}}{1+\frac{g_1 \omega_0}{g_0 \omega_1}e^{-E_1/T}} \; . \label{eq:StatSol}
\end{equation}
In general, the degeneracies $g_0$ and $g_1$ depend upon the landscape complexity, which increases as \cite{mezard2012glasses, wales2000energy, stillinger1999exponential} $\sim N! \exp(N)$ . Here we assume $g_0=\exp\left(N\log(\Omega_0) \right)$ and $g_1=\exp\left(N\log(\Omega_1) \right)$. Therefore, in the thermodynamic limit, when $T<T_c$ then $p_0(T)=0$, while when $T>T_c$ then $p_0(T)=1$, where $T_c$ is the first order transition temperature and is defined by the equality \footnote{Here and on, when taken the thermodynamic limit we will assume $\Omega_1/\Omega_0>1$.}
\begin{equation}
T_c=\frac{\epsilon}{\log\left(\Omega_1/\Omega_0 \right)} \; .
\end{equation}



In equilibrium, the system at $T<T_c$ is in the crystalline state while when $T>T_c$ it represents the liquid. When the system experiences a quench, the system may be arrested in metastable states. This will be presented in the following section.



\section{Cooling speed and residual population} \label{sec:quench}

Let us study our model under cooling. In that case, a cooling protocol, i.e. the temperature 
as a function of the time $T(t)$ needs to be specified. Experimentally, a linear cooling is usually used. For obtaining analytical results, an hyperbolic quench is more appropiate. Both coolings produce similar results, except for the size of the glass transition region, associated with the boundary layer of the differential equation \cite{langer1989entropy,naumis}. 
For the hyperbolic quench, $T(t)=T_0/(1+Rt)$ where $T_0$ is the initial temperature at which the system is in thermal equilibrium and $R$ is the cooling rate. In particular, we are interested in the system's dependence with $\omega_1$ when a rapid quench is applied. 

Notice that care must be taken with our notation. Here  $T=T(t)$, and as a result, the population described by Eq. (\ref{master equation}) will be denoted at times by $p(T)$, which should not be confused with the equilibrium probability $p_0(T)$.   
Having said this, let us write Eq. (\ref{master equation}) as follows:
\begin{eqnarray}\label{eq:masterequationRear}
\dot{p}(t)&=&-\frac{\omega_c}{2\pi \gamma}\left(\omega_1 g_0 e^{-\frac{V}{T(t)}}+\omega_0 g_1 e^{-\frac{(V+E_1)}{T(t)}} \right) p(t) \nonumber \\
&&+ \frac{\omega_0 \omega_c}{2\pi \gamma} g_1 e^{-\frac{V+E_1}{T(t)}} \,.
\end{eqnarray}

The solution to this first order non-homogeneous ordinary differential equation is obtained in a straightforward manner yielding the following:
\begin{eqnarray}
p(t)&=& \exp\left( \frac{\omega_c T_0 e^{-\frac{V}{T(t)}}}{2\pi \gamma R} \left( \frac{\omega_1 g_0}{V}  + \frac{\omega_0 g_1 e^{-\frac{E_1}{T(t)}} }{V+E_1}  \right) \right) \nonumber \\ 
 && \left(p(\infty) - \int_t^\infty dt' \frac{\omega_0 \omega_c}{2\pi \gamma} g_1 e^{ -\left(V+E_1 \right)/T(t')} \right. \times \label{eq:p(t)} \\ 
&  &\left.  \exp\left(-\frac{\omega_c T_0 e^{-\frac{V}{T(t)}}}{2\pi \gamma R} \left( \frac{\omega_1 g_0}{V}  + \frac{\omega_0 g_1 e^{-\frac{E_1}{T(t)}}}{V+E_1}  \right) \right) \right) \nonumber \; .
\end{eqnarray}

\begin{figure}[!ht]
\centering
\begin{subfigure}[b]{0.5\textwidth}
\caption{}
\includegraphics[width=3.37in]{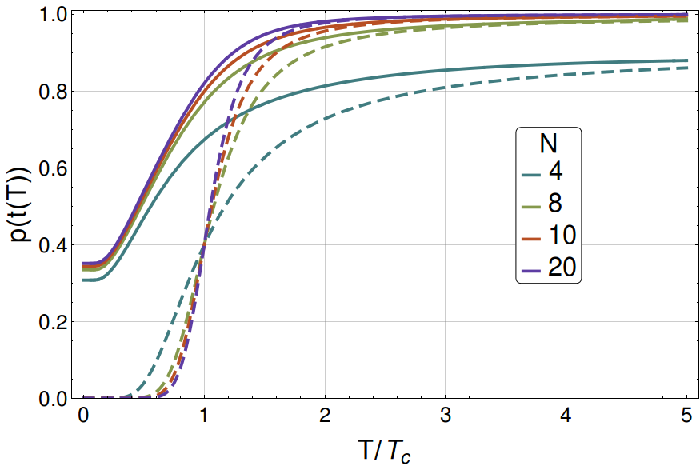}
\end{subfigure}

\begin{subfigure}[b]{0.5\textwidth}
\caption{}
\includegraphics[width=3.37in]{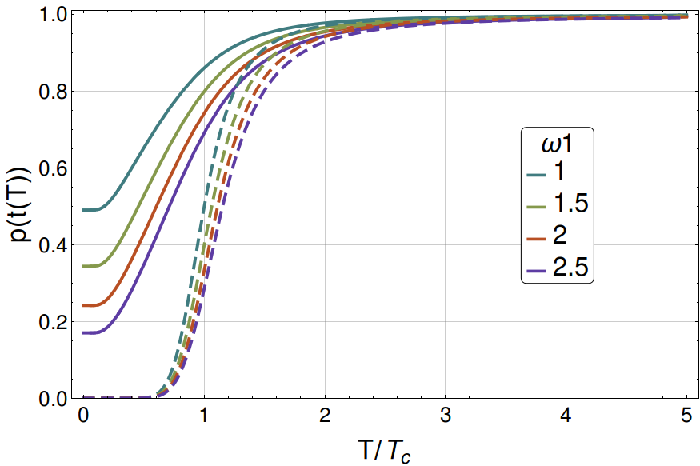}
\end{subfigure}
\caption{\footnotesize{Temperature dependent distribution in equilibrium (dashed lines) and under fast cooling (continuous lines). 
Given the cooling protocol, the system may be arrested in a metastable state. In (a), the size of the system is 
changed for a given cooling ratio. The parameters were 
fixed at \textbf{(a)} $V=1,\; \gamma=1,\; R=1,\; \omega_c=1,\;\omega_0=1,\; \omega_1=1,\; \epsilon=1,\Omega_0=1,\; \Omega_1=2$ . In 
panel \textbf{(b)}, the size and cooling rate is fixed, while the oscillation frequency of the metastable energy basin is modified. Observe 
that the glass forming ability increases as the oscillation frequency $\omega_1$ of the metastable states goes to zero. The reason 
is that the system probes the energy barrier less frequently. 
The parameters are $V=1,\; \gamma=1,\; R=1,\;\omega_c=1,\; \omega_0=1,\; N=8,\; \epsilon=1,\Omega_0=1,\; \Omega_1=2$.}} \label{fig:HighDampingpont}
\end{figure}

Now, in order to find the residual  population $p(\infty)$ corresponding to $t \rightarrow \infty$ which gives the probability of arresting the system in the metastable states as $T\rightarrow 0$, we first assume that the system is initially in thermal equilibrium at a temperature $T_0$ such that $T_0>T_c$. Hence we write:
\begin{eqnarray}
p(\infty)&=& p_0(T_0) \exp\left( -\frac{\omega_c T_0 e^{-\frac{V}{T(0)}}}{2\pi \gamma R} \left( \frac{\omega_1 g_0 }{V}  + \frac{\omega_0 g_1 e^{-\frac{E_1}{T(0)}}}{V+E_1}  \right) \right) \nonumber \\ 
&&+\int_0^\infty dt' \frac{\omega_0 \omega_c}{2\pi \gamma} g_1 e^{ -\left(V+E_1 \right)/T(t')} \times \label{eq:p(infty)} \\ \nonumber
&& \exp\left(-\frac{\omega_c T_0 e^{-\frac{V}{T(t)}}}{2\pi \gamma R} \left( \frac{\omega_1 g_0 }{V}  + \frac{\omega_0 g_1 e^{-\frac{E_1}{T(t)}} }{V+E_1}  \right) \right) \; .
\end{eqnarray}

\begin{figure}[!ht]
\centering
\begin{subfigure}[b]{0.5\textwidth}
\caption{}
\includegraphics[width=3.37in]{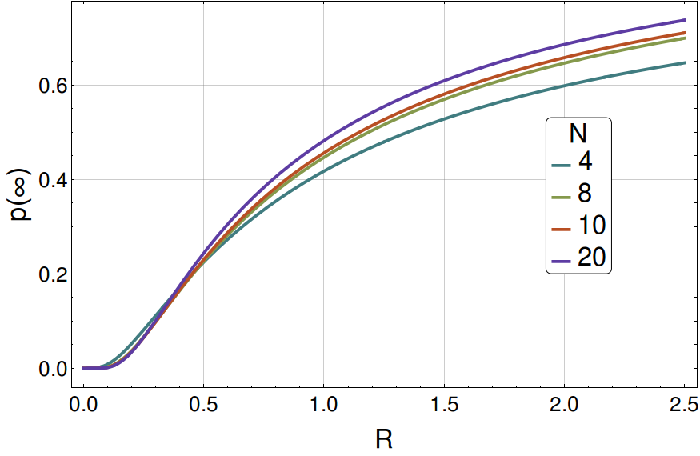}
\end{subfigure}

\begin{subfigure}[b]{0.5\textwidth}
\caption{}
\includegraphics[width=3.37in]{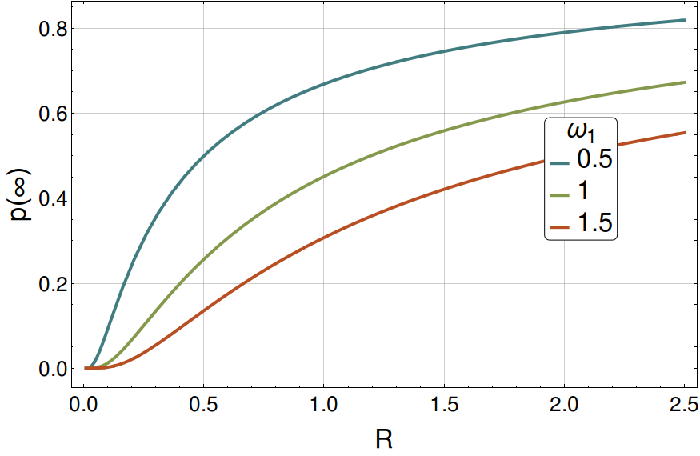}
\end{subfigure}
\caption{\footnotesize{Final state quenched distribution as a function of the cooling rate as obtained from Eq. (\ref{eq:p(infty)}). In panel (a), the size of the system is changed. The parameters were fixed at \textbf{(a)} $V=1,\; \gamma=1,\; T_0=T_c,\; \omega_c=1,\;\omega_0=1,\; \omega_1=1.1,\; \epsilon=1,\Omega_0=1,\; \Omega_1=2$. In panel \textbf{(b)}, the oscillation frequency of the metastable energy basin is modified. Observe how as the frequency $\omega_1 \rightarrow 0$, the glass forming ability increases for a given cooling rate $R$. Again, the reason is the decreasing probing of the energy barrier. The parameters are $V=1,\; \gamma=1,\; T_0=T_c,\;\omega_c=1,\; \omega_0=1,\; N=4,\; \epsilon=1,\; \Omega_0=1,\; \Omega_1=2$.}} \label{fig:HighDampingpinf}
\end{figure}

In Fig. \ref{fig:HighDampingpont} we have plotted $p$ given by Eqs. (\ref{eq:StatSol}) and (\ref{eq:p(t)}) as function of $T$, while in Fig. \ref{fig:HighDampingpinf} we have plotted $p(\infty)$ given by Eq. (\ref{eq:p(infty)}). Notice, from the lower panels in both figures, how the residual population increases as $\omega_1$ decreases, i.e., as the metastable wells become broader. This agrees with the first passage time of a non-drifting Brownian particle (see \cite{redner2001guide} for instance). Notice that as $\omega_1$ tends to zero, the well becomes flat. Hence, the system can be thought of as a one-dimensional free Brownian particle. In this scenario, at long times, the first passage time distribution goes as $\sim t^{-1/2}$. Thus, the mean first passage time does not converge, which means the particle takes an infinite time in going from the metastable state to the ground state. This is what the lower panels in Figs. \ref{fig:HighDampingpont} and \ref{fig:HighDampingpinf} are suggesting.

As previously mentioned, prior to the glass transition, the characteristic relaxation time increases. When this time is of the order of the observation time, then the supercooled liquid is not able to maintain in equilibrium and the glass is formed. This characteristic relaxation time is obtained in the following section.

\section{Characteristic relaxation time} \label{sec:char}
As is well known, glasses appear because the system is not able to relax into the energy minimum. In this simple model we can test this idea in a simple way.
To determine the characteristic relaxation time, let us assume that at any fixed given temperature $T$, the initial condition is $p(t=0)=\rho$, where $0\leq\rho \leq 1$. Now, because of detailed balance, we know that for a {\it fixed temperature} $p(t\rightarrow \infty)=p_0(T)$. Hence, we formally write:
\begin{equation}
\begin{cases}
\dot{p}(t)=-\Gamma_{10}(T)g_0 p(t) + \Gamma_{01} g_1 \left(1-p(t) \right) \\
p(t=0)=\rho \\
p(t\rightarrow \infty)= p_0(T) 
\end{cases} \qquad .
\end{equation}
By simple inspection, one is able to write the solution,
\begin{equation}
p(t)=p_0(T)+\left(\rho - p_0(T) \right) e^{-t/\tau} \; .
\end{equation}
where the characteristic relaxation time $\tau$ is:
\begin{equation}
\tau=\frac{1}{\Gamma_{10}g_0 + \Gamma_{01} g_1}=\frac{2\pi \gamma}{\omega_c \omega_1} e^{(V/T+F/T)} \; , \label{eq:relaxTime}
\end{equation}
where $F$ is the free energy, i.e., 
\begin{equation}
F=-T\log\left(g_0+g_1 \omega_0 \exp\left(-E_1/T \right)/\omega_1 \right).
\end{equation}

Notice that in Eq. (\ref{eq:relaxTime}), $\tau$ is proportional to the oscillation period multiplied by the inverse conditional probability $\mathcal{P}(V+E_1 \vert E_1)$. Before continuing, let us stress the following. When $\gamma \gg 1$ means the system is strongly coupled with the heat bath, thus it dissipates energy at a very high rate. On the contrary, when $\gamma \rightarrow 0$ the system is weakly coupled with the heat bath, which are responsible for the fluctuations in the system which in turn are responsible for the barrier crossing. However, Eq. (\ref{eq:relaxTime}) does not apply for the latter case. Instead, one may use Kramers' low damping regime escape time (see for instance \cite{coffey2004applications, zwanzig2001nonequilibrium, mel1986theory}) 


In the thermodynamical limit, the characteristic relaxation time below the critical temperature goes as $2\pi \gamma/\left( \omega_c \omega_1 \Omega_0^N \right) \exp\left(V/T \right)$.
Despite stating the obvious, notice that as $\omega_1$ decreases, the characteristic relaxation time increases. This is consistent with our previous results. As stated earlier, when $\omega_1$ tends to zero, the energy landscape changes in such a way that the available phase space increases. Hence, it takes longer for the particle to visit the "summit" or probe the energy barrier . Therefore, the characteristic relaxation time increases as $\omega_1$ tends to zero. In this limit, such degree of freedom becomes a floppy mode. As we discuss in the following section, this affects the critical cooling rate for glass formation.


\section{Thermodynamics limits and critical cooling rate} \label{sec:therm}
Let us first consider the thermodynamical limit $N\rightarrow \infty$ for expressions $p(\infty)$ and $p(t)$. In this scenario, from Eq. (\ref{eq:p(infty)}) we obtain for $p(\infty)$ 
the following expression:
\begin{equation}
p(\infty)=\exp \left(-\frac{\Gamma_{10}(T_c) T_0}{RV} \right) \; .
\end{equation}

Arguing the same way, one obtains from Eq. (\ref{eq:p(t)}) for $p(t)$ the following result:
\begin{eqnarray}
p\left( T(t)\right)=
\begin{cases}
1, \qquad T(t)\geq T_c  \\
\exp\left(\frac{\Gamma \left(T(t)\right)  T_0}{ R V} \left(1-e^{V\frac{T_c-T(t)}{T_c T(t)}} \right) \right), \; \; T(t)\leq T_c  ,
\end{cases} \nonumber 
\end{eqnarray}
We define the critical cooling rate as the cooling rate for which the residual population has an inflection point, to obtain the following relation:
\begin{equation}
R_{\text{crit}}=\frac{\omega_1 \omega_c T_0}{4\pi \gamma V} e^{-V/T_c} \; . \label{eq:RCrit}
\end{equation}
This equation relates the cooling rate with the short-time dynamics in the model which is one the main result of this work. In the following section, we will discuss its
properties and validity.

\section{Discussion} \label{sec:Disc}

In the previous sections, we found that Eqs. (\ref{eq:relaxTime}) and (\ref{eq:RCrit}) provide a link between long and short-time dynamics for a simple landscape. Let us
know discuss some important points concerning its application in real systems.

The first is to observe that in Eq. (\ref{eq:RCrit}), $R_{\text{crit}}$ is linear on  $\omega_1$. As $\omega_1 \rightarrow 0$, the relaxation time grows. 
The reason is simple to understand. As the energy wells flattens, the time spent by the system 
close to the dividing energy barrier goes to zero and the probability of escape decreases. In other words, the frequency of
oscillation is roughly the inverse of the time between collisions with the energy barrier. Up to our knowledge this observation has not been taken into
account for the dynamical analysis of glasses. 

We believe this issue has been overlooked due to other effects that also modify the relaxation. All of them play a role. Here we isolated one
of the ingredients, the basin oscillation frequency. Other ingredients are the correlation between the barrier heights and basin oscillation frequency, as well
as the existence of a distribution of basins \cite{sastry, doliwa2003energy, broderix2000energy, Wales2} .

Such effects have been also found from empirical arguments in rigidity 
theory of glasses \cite{naumis2006variation}. A simple and intuitive way to understand this is as follows.
According to Dyre \cite{dyre2006col}, the energy barriers are related with the mean-square displacement $\langle u^{2} \rangle$ by,
\begin{equation}
\Delta E =\lambda_1 k_{B}T\frac{a^{2}}{\langle u^{2} \rangle} \; ,
\end{equation}
with $a$ being the lattice parameter and $\lambda_1$ a factor of order unity. But the mean-square displacement in a basin can be written as \cite{naumis2006variation, flores},  
\begin{equation}
 \langle u^{2} \rangle=\frac{3T}{\langle M \rangle} \int_{0}^{\infty}   \frac{\rho(\omega)}{\omega^{2}}d\omega \; ,
\end{equation}
where $\rho (\omega)$ is the density of vibrational states. Observe that the previous equation holds for the supercooled liquid close to the glass transition as long 
as one performs its computation in a distribution of basins and by using a cut-off for small frequencies \cite{flores2012mean}.
Combining the previous equations we obtain an estimate of the energy barriers,
\begin{equation}
 \Delta E =\lambda_1 \frac{a^{2} \langle M \rangle}{3 \int_{0}^{\infty}   \frac{\rho(\omega)}{\omega^{2}}d\omega} \; .
\end{equation}

Assuming the model presented in Ref. \cite{naumis2006variation} for the DOS of floppy systems, i.e.,
\begin{equation}
g(\omega)=(1-f)g_R(\omega)+f\delta(\omega-\omega_f) \; ,
\end{equation}
we obtain the following:
\begin{eqnarray}
\Delta E &=&\frac{\lambda_1 a^2 \langle M \rangle}{3} \left[ \left(1-f \right)\frac{V \omega_D}{2 \pi^2 c^3 3N} + \frac{f}{\omega_f^2} \right]^{-1} \nonumber \\
&\approx& \frac{\lambda_1 a^2 \langle M \rangle \omega_f^2}{3 f} \left( 1-\frac{V \omega_D \omega_f^2}{2 \pi^2 c^3 3N} \frac{\left(1-f \right)}{f}  \right) \; .
\end{eqnarray}
Thus, energy barriers decrease when the oscillation frequency goes to zero. Interestingly, this suggest a feedback mechanism
on energy barriers and floppy modes, as has been made in the temperature-dependent constraint theory \cite{Gupta2009}.

From the previous considerations, in our model we may assume a more general form of Eq. (\ref{eq:RCrit}), by explicitly taking into account the correlation between
$V$ and $\omega_1$,
\begin{equation}
R_{\text{crit}}=\frac{\omega_1 \omega_c T_0}{4\pi \gamma V(\omega_1)} e^{-V(\omega_1)/T_c} \; . 
\end{equation}

\begin{figure}[hbtp]
\centering
\begin{subfigure}[b]{0.5\textwidth}
\caption{}
\includegraphics[width=3.37in]{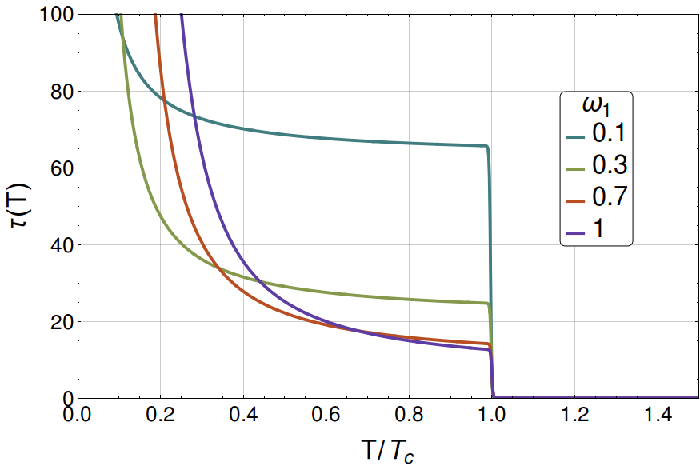}
\end{subfigure}

\begin{subfigure}[b]{0.5\textwidth}
\caption{}
\includegraphics[width=3.37in]{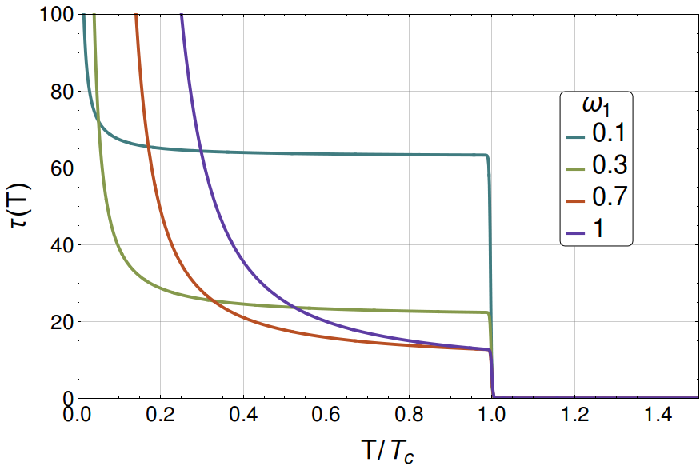}
\end{subfigure}
\caption{\footnotesize{Characteristic relaxation times as a function of the temperature using different oscillation frequencies of the metastable state, as obtained from Eq. (\ref{eq:relaxTime}) considering $V(\omega_1)\sim \omega_1^{1+q}$. In panel \textbf{(a)}, the energy barrier has $q=0.2$ while in panel \textbf{(b)} $q=1$.  The parameters were fixed at: $N=1000,\; \gamma=1,\; \omega_0=1,\; \omega_c=1,\; \epsilon=1,\Omega_0=1,\; \Omega_1=2$.}} \label{fig:CharRelTimeVproptoOmega}
\end{figure}

Here $V(\omega_1)$ denotes such possible correlation. Obviously, its actual form depends upon the particular potential form. However, in its more crude approximation one can extend 
the harmonic approximation around the closer metastable minima to estimate the height of the barrier. For this harmonic approximation, the transition barrier, which goes as $\sim \omega_1^2$, is an overestimation proportional to the separation between the barrier and the minimum, as discussed by Dyre \cite{dyre2006col}. In the case of a quartic double well, the transition barrier also goes as $\sim \omega_1^2$.  For the sake of the argument, let us assume in general that the energy landscape can be written in such a way that $V(\omega_1)\sim \omega_1^{1+q}$ where $q>0$. Then the critical cooling rate is now,
\begin{equation}
R_{\text{crit}}  \sim \frac{1}{\omega_1^{q}} \exp(-\omega_1^{1+q}/T_c) \; .
\label{RcritCorre}
\end{equation}
Notice that this last expression diverges as $\omega_1$ tends to zero. Also, in Fig. \ref{fig:CharRelTimeVproptoOmega}, we have plotted the characteristic relaxation time taking into account this dependence between $V$ and $\omega_1$ for different values of $q$. Notice how there always exists a temperature in which the characteristic relaxation time for broader wells is always smaller than that for narrower wells,
although this can happen at a temperature much lower than $T_c$.
In any case, from Eq. (\ref{RcritCorre}) is clear that {\it the long-time relaxation
depends on the short-time relaxation factor $\omega_1$}. However, its actual functional form depends upon the correlation between energy barrier and short-time oscillation frequency.

Finally, it is worthwhile to mention that the model reproduces the experimentally observed logarithm change with the cooling rate of the glass transition temperature $T_g$. First we observe that $T_g$ is the temperature for which the specific heat has a peak as a function of $T$. From this, one can adapt the approach used by Trachenko \textit{et al.} \cite{trachenko2011heat}, to put the relaxation time Eq. (\ref{eq:relaxTime}) at $T_g$ into the cooling protocol $T=T(t)$, to obtain that, 
\begin{equation}
T_g=\frac{V}{\log\left(\frac{\overline{\Delta T}}{T_0}\right)- \log\left(\frac{2 \pi \gamma R}{\omega_c \omega_1} \right)} \; . \label{eq:Tg}
\end{equation}
Here $\overline{\Delta T}=T_0^2\left( T_1-T_2 \right)/T_1T_2$ is defined as a reduced temperature range between the two temperatures $T_1$ and $T_2(<T_1)$ related by the glass transition relaxation time when the hyperbolic quench is applied. Notice, from Eq. (\ref{eq:Tg}), that $2 \pi \gamma/ \omega_c \omega_1$ is the Debye vibrational period, which is of the order of flow time events \cite{trachenko2011heat}.

\section{Conclusions} \label{sec:conclusions}

In this article we used the Kramers formula to understand the role of a metastable state harmonic oscillation frequency  in a simple model of glass relaxation. The Kramers formula is an improvement of the Arrhenius relaxation formula for the escape time in a well. The effect of such
frequency (related with the curvature of the energy landscape basin) is to decrease the frequency of collisions with the energy barriers. In fact, here we showed that the short time dynamics always enters as a linear factor in the relaxation time, which multiplies the already well known energy barrier exponential factor. This implies that short-time dynamics is important for long-time relaxation. Furthermore, since the relaxation time contains an exponential factor of the energy barrier, it can depend upon the correlation between energy barriers and short-time dynamics for many realistic potentials, the actual functional form of the relaxation time depends on the short-time dynamics. Thus, our work highlights an important feature that has not been taken into account for glass relaxation models, although the Gupta-Mauro temperature-constraint model implicitly incorporates such feature \cite{Gupta2009}.

\section{Acknowledgments} \label{sec:Ack}

The authors would like to thank the referees for constructive criticism on the manuscript. This work was supported by DGAPA-UNAM project IN102717. J.Q.T.M. acknowledges a doctoral fellowship from CONACyT.

\bibliography{mybib}

\begin{thebibliography}{66}%
\makeatletter
\providecommand \@ifxundefined [1]{%
 \@ifx{#1\undefined}
}%
\providecommand \@ifnum [1]{%
 \ifnum #1\expandafter \@firstoftwo
 \else \expandafter \@secondoftwo
 \fi
}%
\providecommand \@ifx [1]{%
 \ifx #1\expandafter \@firstoftwo
 \else \expandafter \@secondoftwo
 \fi
}%
\providecommand \natexlab [1]{#1}%
\providecommand \enquote  [1]{``#1''}%
\providecommand \bibnamefont  [1]{#1}%
\providecommand \bibfnamefont [1]{#1}%
\providecommand \citenamefont [1]{#1}%
\providecommand \href@noop [0]{\@secondoftwo}%
\providecommand \href [0]{\begingroup \@sanitize@url \@href}%
\providecommand \@href[1]{\@@startlink{#1}\@@href}%
\providecommand \@@href[1]{\endgroup#1\@@endlink}%
\providecommand \@sanitize@url [0]{\catcode `\\12\catcode `\$12\catcode
  `\&12\catcode `\#12\catcode `\^12\catcode `\_12\catcode `\%12\relax}%
\providecommand \@@startlink[1]{}%
\providecommand \@@endlink[0]{}%
\providecommand \url  [0]{\begingroup\@sanitize@url \@url }%
\providecommand \@url [1]{\endgroup\@href {#1}{\urlprefix }}%
\providecommand \urlprefix  [0]{URL }%
\providecommand \Eprint [0]{\href }%
\providecommand \doibase [0]{http://dx.doi.org/}%
\providecommand \selectlanguage [0]{\@gobble}%
\providecommand \bibinfo  [0]{\@secondoftwo}%
\providecommand \bibfield  [0]{\@secondoftwo}%
\providecommand \translation [1]{[#1]}%
\providecommand \BibitemOpen [0]{}%
\providecommand \bibitemStop [0]{}%
\providecommand \bibitemNoStop [0]{.\EOS\space}%
\providecommand \EOS [0]{\spacefactor3000\relax}%
\providecommand \BibitemShut  [1]{\csname bibitem#1\endcsname}%
\let\auto@bib@innerbib\@empty
\bibitem [{\citenamefont {Dyre}(2006)}]{dyre2006col}%
  \BibitemOpen
  \bibfield  {author} {\bibinfo {author} {\bibfnamefont {J.~C.}\ \bibnamefont
  {Dyre}},\ }\href@noop {} {\bibfield  {journal} {\bibinfo  {journal} {Reviews
  of modern physics}\ }\textbf {\bibinfo {volume} {78}},\ \bibinfo {pages}
  {953} (\bibinfo {year} {2006})}\BibitemShut {NoStop}%
\bibitem [{\citenamefont {Ngai}(2000)}]{ngai}%
  \BibitemOpen
  \bibfield  {author} {\bibinfo {author} {\bibfnamefont {K.}~\bibnamefont
  {Ngai}},\ }\href@noop {} {\bibfield  {journal} {\bibinfo  {journal} {Journal
  of Physics: Condensed Matter}\ }\textbf {\bibinfo {volume} {12}},\ \bibinfo
  {pages} {6437} (\bibinfo {year} {2000})}\BibitemShut {NoStop}%
\bibitem [{\citenamefont {Trachenko}\ and\ \citenamefont
  {Brazhkin}(2011)}]{trachenko2011heat}%
  \BibitemOpen
  \bibfield  {author} {\bibinfo {author} {\bibfnamefont {K.}~\bibnamefont
  {Trachenko}}\ and\ \bibinfo {author} {\bibfnamefont {V.}~\bibnamefont
  {Brazhkin}},\ }\href@noop {} {\bibfield  {journal} {\bibinfo  {journal}
  {Physical Review B}\ }\textbf {\bibinfo {volume} {83}},\ \bibinfo {pages}
  {014201} (\bibinfo {year} {2011})}\BibitemShut {NoStop}%
\bibitem [{\citenamefont {Dyre}(1987)}]{Dyre1987}%
  \BibitemOpen
  \bibfield  {author} {\bibinfo {author} {\bibfnamefont {J.~C.}\ \bibnamefont
  {Dyre}},\ }\href@noop {} {\bibfield  {journal} {\bibinfo  {journal} {Physical
  Review Letters}\ }\textbf {\bibinfo {volume} {58}},\ \bibinfo {pages} {792}
  (\bibinfo {year} {1987})}\BibitemShut {NoStop}%
\bibitem [{\citenamefont {Dyre}(1995)}]{Dyre1995}%
  \BibitemOpen
  \bibfield  {author} {\bibinfo {author} {\bibfnamefont {J.~C.}\ \bibnamefont
  {Dyre}},\ }\href@noop {} {\bibfield  {journal} {\bibinfo  {journal} {Physical
  Review B}\ }\textbf {\bibinfo {volume} {51}},\ \bibinfo {pages} {12276}
  (\bibinfo {year} {1995})}\BibitemShut {NoStop}%
\bibitem [{\citenamefont {Langer}\ and\ \citenamefont
  {Sethna}(1988)}]{langer1988entropy}%
  \BibitemOpen
  \bibfield  {author} {\bibinfo {author} {\bibfnamefont {S.~A.}\ \bibnamefont
  {Langer}}\ and\ \bibinfo {author} {\bibfnamefont {J.~P.}\ \bibnamefont
  {Sethna}},\ }\href@noop {} {\bibfield  {journal} {\bibinfo  {journal}
  {Physical review letters}\ }\textbf {\bibinfo {volume} {61}},\ \bibinfo
  {pages} {570} (\bibinfo {year} {1988})}\BibitemShut {NoStop}%
\bibitem [{\citenamefont {Naumis}(2006)}]{naumis2006variation}%
  \BibitemOpen
  \bibfield  {author} {\bibinfo {author} {\bibfnamefont {G.~G.}\ \bibnamefont
  {Naumis}},\ }\href@noop {} {\bibfield  {journal} {\bibinfo  {journal}
  {Physical Review B}\ }\textbf {\bibinfo {volume} {73}},\ \bibinfo {pages}
  {172202} (\bibinfo {year} {2006})}\BibitemShut {NoStop}%
\bibitem [{\citenamefont {Phillips}(1996)}]{phillips1996stretched}%
  \BibitemOpen
  \bibfield  {author} {\bibinfo {author} {\bibfnamefont {J.}~\bibnamefont
  {Phillips}},\ }\href@noop {} {\bibfield  {journal} {\bibinfo  {journal}
  {Reports on Progress in Physics}\ }\textbf {\bibinfo {volume} {59}},\
  \bibinfo {pages} {1133} (\bibinfo {year} {1996})}\BibitemShut {NoStop}%
\bibitem [{\citenamefont {Naumis}\ and\ \citenamefont
  {Kerner}(1998)}]{naumis1998stochastic}%
  \BibitemOpen
  \bibfield  {author} {\bibinfo {author} {\bibfnamefont {G.~G.}\ \bibnamefont
  {Naumis}}\ and\ \bibinfo {author} {\bibfnamefont {R.}~\bibnamefont
  {Kerner}},\ }\href@noop {} {\bibfield  {journal} {\bibinfo  {journal}
  {Journal of non-crystalline solids}\ }\textbf {\bibinfo {volume} {231}},\
  \bibinfo {pages} {111} (\bibinfo {year} {1998})}\BibitemShut {NoStop}%
\bibitem [{\citenamefont {Micoulaut}\ and\ \citenamefont
  {Naumis}(1999)}]{micoulaut1999glass}%
  \BibitemOpen
  \bibfield  {author} {\bibinfo {author} {\bibfnamefont {M.}~\bibnamefont
  {Micoulaut}}\ and\ \bibinfo {author} {\bibfnamefont {G.}~\bibnamefont
  {Naumis}},\ }\href@noop {} {\bibfield  {journal} {\bibinfo  {journal} {EPL
  (Europhysics Letters)}\ }\textbf {\bibinfo {volume} {47}},\ \bibinfo {pages}
  {568} (\bibinfo {year} {1999})}\BibitemShut {NoStop}%
\bibitem [{\citenamefont {Kerner}\ and\ \citenamefont
  {Naumis}(2000)}]{kerner2000stochastic}%
  \BibitemOpen
  \bibfield  {author} {\bibinfo {author} {\bibfnamefont {R.}~\bibnamefont
  {Kerner}}\ and\ \bibinfo {author} {\bibfnamefont {G.~G.}\ \bibnamefont
  {Naumis}},\ }\href@noop {} {\bibfield  {journal} {\bibinfo  {journal}
  {Journal of Physics: Condensed Matter}\ }\textbf {\bibinfo {volume} {12}},\
  \bibinfo {pages} {1641} (\bibinfo {year} {2000})}\BibitemShut {NoStop}%
\bibitem [{\citenamefont {Smedskjaer}, \citenamefont {Mauro},\ and\
  \citenamefont {Yue}(2010)}]{Mauro1}%
  \BibitemOpen
  \bibfield  {author} {\bibinfo {author} {\bibfnamefont {M.~M.}\ \bibnamefont
  {Smedskjaer}}, \bibinfo {author} {\bibfnamefont {J.~C.}\ \bibnamefont
  {Mauro}}, \ and\ \bibinfo {author} {\bibfnamefont {Y.}~\bibnamefont {Yue}},\
  }\href@noop {} {\bibfield  {journal} {\bibinfo  {journal} {Physical review
  letters}\ }\textbf {\bibinfo {volume} {105}},\ \bibinfo {pages} {115503}
  (\bibinfo {year} {2010})}\BibitemShut {NoStop}%
\bibitem [{\citenamefont {Mauro}, \citenamefont {Allan},\ and\ \citenamefont
  {Potuzak}(2009)}]{Mauro2}%
  \BibitemOpen
  \bibfield  {author} {\bibinfo {author} {\bibfnamefont {J.~C.}\ \bibnamefont
  {Mauro}}, \bibinfo {author} {\bibfnamefont {D.~C.}\ \bibnamefont {Allan}}, \
  and\ \bibinfo {author} {\bibfnamefont {M.}~\bibnamefont {Potuzak}},\
  }\href@noop {} {\bibfield  {journal} {\bibinfo  {journal} {Physical Review
  B}\ }\textbf {\bibinfo {volume} {80}},\ \bibinfo {pages} {094204} (\bibinfo
  {year} {2009})}\BibitemShut {NoStop}%
\bibitem [{\citenamefont {Mauro}\ \emph {et~al.}(2009)\citenamefont {Mauro},
  \citenamefont {Yue}, \citenamefont {Ellison}, \citenamefont {Gupta},\ and\
  \citenamefont {Allan}}]{mauro3PNAS}%
  \BibitemOpen
  \bibfield  {author} {\bibinfo {author} {\bibfnamefont {J.~C.}\ \bibnamefont
  {Mauro}}, \bibinfo {author} {\bibfnamefont {Y.}~\bibnamefont {Yue}}, \bibinfo
  {author} {\bibfnamefont {A.~J.}\ \bibnamefont {Ellison}}, \bibinfo {author}
  {\bibfnamefont {P.~K.}\ \bibnamefont {Gupta}}, \ and\ \bibinfo {author}
  {\bibfnamefont {D.~C.}\ \bibnamefont {Allan}},\ }\href@noop {} {\bibfield
  {journal} {\bibinfo  {journal} {Proceedings of the National Academy of
  Sciences}\ }\textbf {\bibinfo {volume} {106}},\ \bibinfo {pages} {19780}
  (\bibinfo {year} {2009})}\BibitemShut {NoStop}%
\bibitem [{\citenamefont {Debenedetti}(1996)}]{debenedetti1996metastable}%
  \BibitemOpen
  \bibfield  {author} {\bibinfo {author} {\bibfnamefont {P.~G.}\ \bibnamefont
  {Debenedetti}},\ }\href@noop {} {\emph {\bibinfo {title} {Metastable liquids:
  concepts and principles}}}\ (\bibinfo  {publisher} {Princeton University
  Press},\ \bibinfo {year} {1996})\BibitemShut {NoStop}%
\bibitem [{\citenamefont {Debenedetti}\ and\ \citenamefont
  {Stillinger}(2001)}]{debenedetti2001supercooled}%
  \BibitemOpen
  \bibfield  {author} {\bibinfo {author} {\bibfnamefont {P.~G.}\ \bibnamefont
  {Debenedetti}}\ and\ \bibinfo {author} {\bibfnamefont {F.~H.}\ \bibnamefont
  {Stillinger}},\ }\href@noop {} {\bibfield  {journal} {\bibinfo  {journal}
  {Nature}\ }\textbf {\bibinfo {volume} {410}},\ \bibinfo {pages} {259}
  (\bibinfo {year} {2001})}\BibitemShut {NoStop}%
\bibitem [{\citenamefont {Stillinger}\ and\ \citenamefont
  {Debenedetti}(2002)}]{stillinger2002energy}%
  \BibitemOpen
  \bibfield  {author} {\bibinfo {author} {\bibfnamefont {F.~H.}\ \bibnamefont
  {Stillinger}}\ and\ \bibinfo {author} {\bibfnamefont {P.~G.}\ \bibnamefont
  {Debenedetti}},\ }\href@noop {} {\bibfield  {journal} {\bibinfo  {journal}
  {The Journal of chemical physics}\ }\textbf {\bibinfo {volume} {116}},\
  \bibinfo {pages} {3353} (\bibinfo {year} {2002})}\BibitemShut {NoStop}%
\bibitem [{\citenamefont {Wang}\ \emph {et~al.}(2001)\citenamefont {Wang},
  \citenamefont {Wells}, \citenamefont {Georgiev}, \citenamefont {Boolchand},
  \citenamefont {Jackson},\ and\ \citenamefont {Micoulaut}}]{wang2001sharp}%
  \BibitemOpen
  \bibfield  {author} {\bibinfo {author} {\bibfnamefont {Y.}~\bibnamefont
  {Wang}}, \bibinfo {author} {\bibfnamefont {J.}~\bibnamefont {Wells}},
  \bibinfo {author} {\bibfnamefont {D.}~\bibnamefont {Georgiev}}, \bibinfo
  {author} {\bibfnamefont {P.}~\bibnamefont {Boolchand}}, \bibinfo {author}
  {\bibfnamefont {K.}~\bibnamefont {Jackson}}, \ and\ \bibinfo {author}
  {\bibfnamefont {M.}~\bibnamefont {Micoulaut}},\ }\href@noop {} {\bibfield
  {journal} {\bibinfo  {journal} {Physical Review Letters}\ }\textbf {\bibinfo
  {volume} {87}},\ \bibinfo {pages} {185503} (\bibinfo {year}
  {2001})}\BibitemShut {NoStop}%
\bibitem [{\citenamefont {Phillips}(1979)}]{phillips1979topology}%
  \BibitemOpen
  \bibfield  {author} {\bibinfo {author} {\bibfnamefont {J.~C.}\ \bibnamefont
  {Phillips}},\ }\href@noop {} {\bibfield  {journal} {\bibinfo  {journal}
  {Journal of Non-Crystalline Solids}\ }\textbf {\bibinfo {volume} {34}},\
  \bibinfo {pages} {153} (\bibinfo {year} {1979})}\BibitemShut {NoStop}%
\bibitem [{\citenamefont {Naumis}(2005)}]{naumis2005energy}%
  \BibitemOpen
  \bibfield  {author} {\bibinfo {author} {\bibfnamefont {G.~G.}\ \bibnamefont
  {Naumis}},\ }\href@noop {} {\bibfield  {journal} {\bibinfo  {journal}
  {Physical Review E}\ }\textbf {\bibinfo {volume} {71}},\ \bibinfo {pages}
  {026114} (\bibinfo {year} {2005})}\BibitemShut {NoStop}%
\bibitem [{\citenamefont {Sastry}(2001)}]{sastry}%
  \BibitemOpen
  \bibfield  {author} {\bibinfo {author} {\bibfnamefont {S.}~\bibnamefont
  {Sastry}},\ }\href@noop {} {\bibfield  {journal} {\bibinfo  {journal}
  {Nature}\ }\textbf {\bibinfo {volume} {409}},\ \bibinfo {pages} {164}
  (\bibinfo {year} {2001})}\BibitemShut {NoStop}%
\bibitem [{\citenamefont {Mezard}\ and\ \citenamefont
  {Parisi}(2012)}]{mezard2012glasses}%
  \BibitemOpen
  \bibfield  {author} {\bibinfo {author} {\bibfnamefont {M.}~\bibnamefont
  {Mezard}}\ and\ \bibinfo {author} {\bibfnamefont {G.}~\bibnamefont
  {Parisi}},\ }\href@noop {} {\emph {\bibinfo {title} {Glasses and replicas}}}\
  (\bibinfo  {publisher} {John Wiley \& Sons},\ \bibinfo {year} {2012})\ pp.\
  \bibinfo {pages} {151--191}\BibitemShut {NoStop}%
\bibitem [{\citenamefont {Gleim}, \citenamefont {Kob},\ and\ \citenamefont
  {Binder}(1998)}]{gleim1998does}%
  \BibitemOpen
  \bibfield  {author} {\bibinfo {author} {\bibfnamefont {T.}~\bibnamefont
  {Gleim}}, \bibinfo {author} {\bibfnamefont {W.}~\bibnamefont {Kob}}, \ and\
  \bibinfo {author} {\bibfnamefont {K.}~\bibnamefont {Binder}},\ }\href@noop {}
  {\bibfield  {journal} {\bibinfo  {journal} {Physical review letters}\
  }\textbf {\bibinfo {volume} {81}},\ \bibinfo {pages} {4404} (\bibinfo {year}
  {1998})}\BibitemShut {NoStop}%
\bibitem [{\citenamefont {Gleim}\ and\ \citenamefont
  {Kob}(2000)}]{gleim2000relaxation}%
  \BibitemOpen
  \bibfield  {author} {\bibinfo {author} {\bibfnamefont {T.}~\bibnamefont
  {Gleim}}\ and\ \bibinfo {author} {\bibfnamefont {W.}~\bibnamefont {Kob}},\
  }\href@noop {} {\bibfield  {journal} {\bibinfo  {journal} {The European
  Physical Journal B-Condensed Matter and Complex Systems}\ }\textbf {\bibinfo
  {volume} {13}},\ \bibinfo {pages} {83} (\bibinfo {year} {2000})}\BibitemShut
  {NoStop}%
\bibitem [{\citenamefont {Goldstein}(1969)}]{goldstein1969viscous}%
  \BibitemOpen
  \bibfield  {author} {\bibinfo {author} {\bibfnamefont {M.}~\bibnamefont
  {Goldstein}},\ }\href@noop {} {\bibfield  {journal} {\bibinfo  {journal} {The
  Journal of Chemical Physics}\ }\textbf {\bibinfo {volume} {51}},\ \bibinfo
  {pages} {3728} (\bibinfo {year} {1969})}\BibitemShut {NoStop}%
\bibitem [{\citenamefont {Adam}\ and\ \citenamefont
  {Gibbs}(1965)}]{adam1965temperature}%
  \BibitemOpen
  \bibfield  {author} {\bibinfo {author} {\bibfnamefont {G.}~\bibnamefont
  {Adam}}\ and\ \bibinfo {author} {\bibfnamefont {J.~H.}\ \bibnamefont
  {Gibbs}},\ }\href@noop {} {\bibfield  {journal} {\bibinfo  {journal} {The
  journal of chemical physics}\ }\textbf {\bibinfo {volume} {43}},\ \bibinfo
  {pages} {139} (\bibinfo {year} {1965})}\BibitemShut {NoStop}%
\bibitem [{\citenamefont {Debenedetti}\ \emph {et~al.}(2001)\citenamefont
  {Debenedetti}, \citenamefont {Truskett}, \citenamefont {Lewis},\ and\
  \citenamefont {Stillinger}}]{debenedetti2001theory}%
  \BibitemOpen
  \bibfield  {author} {\bibinfo {author} {\bibfnamefont {P.~G.}\ \bibnamefont
  {Debenedetti}}, \bibinfo {author} {\bibfnamefont {T.~M.}\ \bibnamefont
  {Truskett}}, \bibinfo {author} {\bibfnamefont {C.~P.}\ \bibnamefont {Lewis}},
  \ and\ \bibinfo {author} {\bibfnamefont {F.~H.}\ \bibnamefont {Stillinger}},\
  }\href@noop {} {\bibfield  {journal} {\bibinfo  {journal} {Advances in
  Chemical Engineering}\ }\textbf {\bibinfo {volume} {28}},\ \bibinfo {pages}
  {21} (\bibinfo {year} {2001})}\BibitemShut {NoStop}%
\bibitem [{\citenamefont {Milchev}\ and\ \citenamefont
  {Avramov}(1983)}]{milchev1983influence}%
  \BibitemOpen
  \bibfield  {author} {\bibinfo {author} {\bibfnamefont {A.}~\bibnamefont
  {Milchev}}\ and\ \bibinfo {author} {\bibfnamefont {I.}~\bibnamefont
  {Avramov}},\ }\href@noop {} {\bibfield  {journal} {\bibinfo  {journal}
  {physica status solidi (b)}\ }\textbf {\bibinfo {volume} {120}},\ \bibinfo
  {pages} {123} (\bibinfo {year} {1983})}\BibitemShut {NoStop}%
\bibitem [{\citenamefont {Avramov}\ and\ \citenamefont
  {Milchev}(1988)}]{avramov1988effect}%
  \BibitemOpen
  \bibfield  {author} {\bibinfo {author} {\bibfnamefont {I.}~\bibnamefont
  {Avramov}}\ and\ \bibinfo {author} {\bibfnamefont {A.}~\bibnamefont
  {Milchev}},\ }\href@noop {} {\bibfield  {journal} {\bibinfo  {journal}
  {Journal of non-crystalline solids}\ }\textbf {\bibinfo {volume} {104}},\
  \bibinfo {pages} {253} (\bibinfo {year} {1988})}\BibitemShut {NoStop}%
\bibitem [{\citenamefont {Stillinger}(1999)}]{stillinger1999exponential}%
  \BibitemOpen
  \bibfield  {author} {\bibinfo {author} {\bibfnamefont {F.~H.}\ \bibnamefont
  {Stillinger}},\ }\href@noop {} {\bibfield  {journal} {\bibinfo  {journal}
  {Physical Review E}\ }\textbf {\bibinfo {volume} {59}},\ \bibinfo {pages}
  {48} (\bibinfo {year} {1999})}\BibitemShut {NoStop}%
\bibitem [{\citenamefont {Widmer-Cooper}\ and\ \citenamefont
  {Harrowell}(2006)}]{widmer2006predicting}%
  \BibitemOpen
  \bibfield  {author} {\bibinfo {author} {\bibfnamefont {A.}~\bibnamefont
  {Widmer-Cooper}}\ and\ \bibinfo {author} {\bibfnamefont {P.}~\bibnamefont
  {Harrowell}},\ }\href@noop {} {\bibfield  {journal} {\bibinfo  {journal}
  {Physical review letters}\ }\textbf {\bibinfo {volume} {96}},\ \bibinfo
  {pages} {185701} (\bibinfo {year} {2006})}\BibitemShut {NoStop}%
\bibitem [{\citenamefont {Faraone}\ \emph {et~al.}(2004)\citenamefont
  {Faraone}, \citenamefont {Liu}, \citenamefont {Mou}, \citenamefont {Yen},\
  and\ \citenamefont {Chen}}]{faraone2004fragile}%
  \BibitemOpen
  \bibfield  {author} {\bibinfo {author} {\bibfnamefont {A.}~\bibnamefont
  {Faraone}}, \bibinfo {author} {\bibfnamefont {L.}~\bibnamefont {Liu}},
  \bibinfo {author} {\bibfnamefont {C.-Y.}\ \bibnamefont {Mou}}, \bibinfo
  {author} {\bibfnamefont {C.-W.}\ \bibnamefont {Yen}}, \ and\ \bibinfo
  {author} {\bibfnamefont {S.-H.}\ \bibnamefont {Chen}},\ }\href@noop {}
  {\bibfield  {journal} {\bibinfo  {journal} {The Journal of chemical physics}\
  }\textbf {\bibinfo {volume} {121}},\ \bibinfo {pages} {10843} (\bibinfo
  {year} {2004})}\BibitemShut {NoStop}%
\bibitem [{\citenamefont {Trachenko}\ and\ \citenamefont
  {Brazhkin}(2015)}]{trachenko2015collective}%
  \BibitemOpen
  \bibfield  {author} {\bibinfo {author} {\bibfnamefont {K.}~\bibnamefont
  {Trachenko}}\ and\ \bibinfo {author} {\bibfnamefont {V.}~\bibnamefont
  {Brazhkin}},\ }\href@noop {} {\bibfield  {journal} {\bibinfo  {journal}
  {Reports on Progress in Physics}\ }\textbf {\bibinfo {volume} {79}},\
  \bibinfo {pages} {016502} (\bibinfo {year} {2015})}\BibitemShut {NoStop}%
\bibitem [{\citenamefont {Dyre}(2016)}]{dyre2016simple}%
  \BibitemOpen
  \bibfield  {author} {\bibinfo {author} {\bibfnamefont {J.~C.}\ \bibnamefont
  {Dyre}},\ }\href@noop {} {\bibfield  {journal} {\bibinfo  {journal} {Journal
  of Physics: Condensed Matter}\ }\textbf {\bibinfo {volume} {28}},\ \bibinfo
  {pages} {323001} (\bibinfo {year} {2016})}\BibitemShut {NoStop}%
\bibitem [{\citenamefont {Doliwa}\ and\ \citenamefont
  {Heuer}(2003)}]{doliwa2003energy}%
  \BibitemOpen
  \bibfield  {author} {\bibinfo {author} {\bibfnamefont {B.}~\bibnamefont
  {Doliwa}}\ and\ \bibinfo {author} {\bibfnamefont {A.}~\bibnamefont {Heuer}},\
  }\href@noop {} {\bibfield  {journal} {\bibinfo  {journal} {Physical Review
  E}\ }\textbf {\bibinfo {volume} {67}},\ \bibinfo {pages} {031506} (\bibinfo
  {year} {2003})}\BibitemShut {NoStop}%
\bibitem [{\citenamefont {Broderix}\ \emph {et~al.}(2000)\citenamefont
  {Broderix}, \citenamefont {Bhattacharya}, \citenamefont {Cavagna},
  \citenamefont {Zippelius},\ and\ \citenamefont
  {Giardina}}]{broderix2000energy}%
  \BibitemOpen
  \bibfield  {author} {\bibinfo {author} {\bibfnamefont {K.}~\bibnamefont
  {Broderix}}, \bibinfo {author} {\bibfnamefont {K.~K.}\ \bibnamefont
  {Bhattacharya}}, \bibinfo {author} {\bibfnamefont {A.}~\bibnamefont
  {Cavagna}}, \bibinfo {author} {\bibfnamefont {A.}~\bibnamefont {Zippelius}},
  \ and\ \bibinfo {author} {\bibfnamefont {I.}~\bibnamefont {Giardina}},\
  }\href@noop {} {\bibfield  {journal} {\bibinfo  {journal} {Physical review
  letters}\ }\textbf {\bibinfo {volume} {85}},\ \bibinfo {pages} {5360}
  (\bibinfo {year} {2000})}\BibitemShut {NoStop}%
\bibitem [{\citenamefont {Huse}\ and\ \citenamefont
  {Fisher}(1986)}]{huse1986residual}%
  \BibitemOpen
  \bibfield  {author} {\bibinfo {author} {\bibfnamefont {D.~A.}\ \bibnamefont
  {Huse}}\ and\ \bibinfo {author} {\bibfnamefont {D.~S.}\ \bibnamefont
  {Fisher}},\ }\href@noop {} {\bibfield  {journal} {\bibinfo  {journal}
  {Physical review letters}\ }\textbf {\bibinfo {volume} {57}},\ \bibinfo
  {pages} {2203} (\bibinfo {year} {1986})}\BibitemShut {NoStop}%
\bibitem [{\citenamefont {Langer}, \citenamefont {Sethna},\ and\ \citenamefont
  {Grannan}(1990)}]{langer1990nonequilibrium}%
  \BibitemOpen
  \bibfield  {author} {\bibinfo {author} {\bibfnamefont {S.~A.}\ \bibnamefont
  {Langer}}, \bibinfo {author} {\bibfnamefont {J.~P.}\ \bibnamefont {Sethna}},
  \ and\ \bibinfo {author} {\bibfnamefont {E.~R.}\ \bibnamefont {Grannan}},\
  }\href@noop {} {\bibfield  {journal} {\bibinfo  {journal} {Physical Review
  B}\ }\textbf {\bibinfo {volume} {41}},\ \bibinfo {pages} {2261} (\bibinfo
  {year} {1990})}\BibitemShut {NoStop}%
\bibitem [{\citenamefont {Brey}\ and\ \citenamefont
  {Prados}(1991)}]{brey1991residual}%
  \BibitemOpen
  \bibfield  {author} {\bibinfo {author} {\bibfnamefont {J.~J.}\ \bibnamefont
  {Brey}}\ and\ \bibinfo {author} {\bibfnamefont {A.}~\bibnamefont {Prados}},\
  }\href@noop {} {\bibfield  {journal} {\bibinfo  {journal} {Physical Review
  B}\ }\textbf {\bibinfo {volume} {43}},\ \bibinfo {pages} {8350} (\bibinfo
  {year} {1991})}\BibitemShut {NoStop}%
\bibitem [{\citenamefont {Dauchot}\ and\ \citenamefont
  {Bertin}(2014)}]{dauchot2014glass}%
  \BibitemOpen
  \bibfield  {author} {\bibinfo {author} {\bibfnamefont {O.}~\bibnamefont
  {Dauchot}}\ and\ \bibinfo {author} {\bibfnamefont {E.}~\bibnamefont
  {Bertin}},\ }\href@noop {} {\bibfield  {journal} {\bibinfo  {journal} {The
  European Physical Journal E}\ }\textbf {\bibinfo {volume} {37}},\ \bibinfo
  {pages} {1} (\bibinfo {year} {2014})}\BibitemShut {NoStop}%
\bibitem [{\citenamefont {Wales}\ \emph {et~al.}(2000)\citenamefont {Wales},
  \citenamefont {Doye}, \citenamefont {Miller}, \citenamefont {Mortenson},\
  and\ \citenamefont {Walsh}}]{wales2000energy}%
  \BibitemOpen
  \bibfield  {author} {\bibinfo {author} {\bibfnamefont {D.~J.}\ \bibnamefont
  {Wales}}, \bibinfo {author} {\bibfnamefont {J.~P.}\ \bibnamefont {Doye}},
  \bibinfo {author} {\bibfnamefont {M.~A.}\ \bibnamefont {Miller}}, \bibinfo
  {author} {\bibfnamefont {P.~N.}\ \bibnamefont {Mortenson}}, \ and\ \bibinfo
  {author} {\bibfnamefont {T.~R.}\ \bibnamefont {Walsh}},\ }\href@noop {}
  {\bibfield  {journal} {\bibinfo  {journal} {Advances in Chemical Physics}\
  }\textbf {\bibinfo {volume} {115}},\ \bibinfo {pages} {1} (\bibinfo {year}
  {2000})}\BibitemShut {NoStop}%
\bibitem [{\citenamefont {Bi}\ \emph {et~al.}(2016)\citenamefont {Bi},
  \citenamefont {Yang}, \citenamefont {Marchetti},\ and\ \citenamefont
  {Manning}}]{manning}%
  \BibitemOpen
  \bibfield  {author} {\bibinfo {author} {\bibfnamefont {D.}~\bibnamefont
  {Bi}}, \bibinfo {author} {\bibfnamefont {X.}~\bibnamefont {Yang}}, \bibinfo
  {author} {\bibfnamefont {M.~C.}\ \bibnamefont {Marchetti}}, \ and\ \bibinfo
  {author} {\bibfnamefont {M.~L.}\ \bibnamefont {Manning}},\ }\href@noop {}
  {\bibfield  {journal} {\bibinfo  {journal} {Physical Review X}\ }\textbf
  {\bibinfo {volume} {6}},\ \bibinfo {pages} {021011} (\bibinfo {year}
  {2016})}\BibitemShut {NoStop}%
\bibitem [{\citenamefont {Daggett}\ and\ \citenamefont
  {Fersht}(2003)}]{daggett2003there}%
  \BibitemOpen
  \bibfield  {author} {\bibinfo {author} {\bibfnamefont {V.}~\bibnamefont
  {Daggett}}\ and\ \bibinfo {author} {\bibfnamefont {A.~R.}\ \bibnamefont
  {Fersht}},\ }\href@noop {} {\bibfield  {journal} {\bibinfo  {journal} {Trends
  in biochemical sciences}\ }\textbf {\bibinfo {volume} {28}},\ \bibinfo
  {pages} {18} (\bibinfo {year} {2003})}\BibitemShut {NoStop}%
\bibitem [{\citenamefont {Thorpe}(1983)}]{thorpe1983continuous}%
  \BibitemOpen
  \bibfield  {author} {\bibinfo {author} {\bibfnamefont {M.}~\bibnamefont
  {Thorpe}},\ }\href@noop {} {\bibfield  {journal} {\bibinfo  {journal}
  {Journal of Non-Crystalline Solids}\ }\textbf {\bibinfo {volume} {57}},\
  \bibinfo {pages} {355} (\bibinfo {year} {1983})}\BibitemShut {NoStop}%
\bibitem [{\citenamefont {Selvanathan}, \citenamefont {Bresser},\ and\
  \citenamefont {Boolchand}(2000)}]{selvanathan2000stiffness}%
  \BibitemOpen
  \bibfield  {author} {\bibinfo {author} {\bibfnamefont {D.}~\bibnamefont
  {Selvanathan}}, \bibinfo {author} {\bibfnamefont {W.}~\bibnamefont
  {Bresser}}, \ and\ \bibinfo {author} {\bibfnamefont {P.}~\bibnamefont
  {Boolchand}},\ }\href@noop {} {\bibfield  {journal} {\bibinfo  {journal}
  {Physical Review B}\ }\textbf {\bibinfo {volume} {61}},\ \bibinfo {pages}
  {15061} (\bibinfo {year} {2000})}\BibitemShut {NoStop}%
\bibitem [{\citenamefont {Huerta}\ and\ \citenamefont {Naumis}(2002)}]{huerta}%
  \BibitemOpen
  \bibfield  {author} {\bibinfo {author} {\bibfnamefont {A.}~\bibnamefont
  {Huerta}}\ and\ \bibinfo {author} {\bibfnamefont {G.}~\bibnamefont
  {Naumis}},\ }\href@noop {} {\bibfield  {journal} {\bibinfo  {journal}
  {Physics Letters A}\ }\textbf {\bibinfo {volume} {299}},\ \bibinfo {pages}
  {660} (\bibinfo {year} {2002})}\BibitemShut {NoStop}%
\bibitem [{\citenamefont {Landauer}\ and\ \citenamefont
  {Swanson}(1961)}]{landauer1961frequency}%
  \BibitemOpen
  \bibfield  {author} {\bibinfo {author} {\bibfnamefont {R.}~\bibnamefont
  {Landauer}}\ and\ \bibinfo {author} {\bibfnamefont {J.}~\bibnamefont
  {Swanson}},\ }\href@noop {} {\bibfield  {journal} {\bibinfo  {journal}
  {Physical Review}\ }\textbf {\bibinfo {volume} {121}},\ \bibinfo {pages}
  {1668} (\bibinfo {year} {1961})}\BibitemShut {NoStop}%
\bibitem [{\citenamefont {Zwanzig}(2001)}]{zwanzig2001nonequilibrium}%
  \BibitemOpen
  \bibfield  {author} {\bibinfo {author} {\bibfnamefont {R.}~\bibnamefont
  {Zwanzig}},\ }\href@noop {} {\emph {\bibinfo {title} {Nonequilibrium
  statistical mechanics}}}\ (\bibinfo  {publisher} {Oxford University Press,
  USA},\ \bibinfo {year} {2001})\BibitemShut {NoStop}%
\bibitem [{\citenamefont {H{\"a}nggi}, \citenamefont {Talkner},\ and\
  \citenamefont {Borkovec}(1990)}]{hanggi}%
  \BibitemOpen
  \bibfield  {author} {\bibinfo {author} {\bibfnamefont {P.}~\bibnamefont
  {H{\"a}nggi}}, \bibinfo {author} {\bibfnamefont {P.}~\bibnamefont {Talkner}},
  \ and\ \bibinfo {author} {\bibfnamefont {M.}~\bibnamefont {Borkovec}},\
  }\href@noop {} {\bibfield  {journal} {\bibinfo  {journal} {Reviews of modern
  physics}\ }\textbf {\bibinfo {volume} {62}},\ \bibinfo {pages} {251}
  (\bibinfo {year} {1990})}\BibitemShut {NoStop}%
\bibitem [{\citenamefont {Coffey}, \citenamefont {Kalmykov},\ and\
  \citenamefont {Waldron}(2004)}]{coffey2004applications}%
  \BibitemOpen
  \bibfield  {author} {\bibinfo {author} {\bibfnamefont {W.}~\bibnamefont
  {Coffey}}, \bibinfo {author} {\bibfnamefont {Y.~P.}\ \bibnamefont
  {Kalmykov}}, \ and\ \bibinfo {author} {\bibfnamefont {J.}~\bibnamefont
  {Waldron}},\ }\href@noop {} {\emph {\bibinfo {title} {With Applications to
  Stochastic Problems in Physics, Chemistry and Electrical Engineering 2nd
  Edition}}}\ (\bibinfo  {publisher} {World Scientific},\ \bibinfo {year}
  {2004})\BibitemShut {NoStop}%
\bibitem [{\citenamefont {Flores-Ruiz}\ and\ \citenamefont
  {Naumis}(2012)}]{flores2012mean}%
  \BibitemOpen
  \bibfield  {author} {\bibinfo {author} {\bibfnamefont {H.~M.}\ \bibnamefont
  {Flores-Ruiz}}\ and\ \bibinfo {author} {\bibfnamefont {G.~G.}\ \bibnamefont
  {Naumis}},\ }\href@noop {} {\bibfield  {journal} {\bibinfo  {journal}
  {Physical Review E}\ }\textbf {\bibinfo {volume} {85}},\ \bibinfo {pages}
  {041503} (\bibinfo {year} {2012})}\BibitemShut {NoStop}%
\bibitem [{\citenamefont {Flores-Ruiz}, \citenamefont {Naumis},\ and\
  \citenamefont {Phillips}(2010)}]{flores}%
  \BibitemOpen
  \bibfield  {author} {\bibinfo {author} {\bibfnamefont {H.~M.}\ \bibnamefont
  {Flores-Ruiz}}, \bibinfo {author} {\bibfnamefont {G.~G.}\ \bibnamefont
  {Naumis}}, \ and\ \bibinfo {author} {\bibfnamefont {J.}~\bibnamefont
  {Phillips}},\ }\href@noop {} {\bibfield  {journal} {\bibinfo  {journal}
  {Physical Review B}\ }\textbf {\bibinfo {volume} {82}},\ \bibinfo {pages}
  {214201} (\bibinfo {year} {2010})}\BibitemShut {NoStop}%
\bibitem [{\citenamefont {Flores-Ruiz}\ and\ \citenamefont
  {Naumis}(2011)}]{flores2011boson}%
  \BibitemOpen
  \bibfield  {author} {\bibinfo {author} {\bibfnamefont {H.~M.}\ \bibnamefont
  {Flores-Ruiz}}\ and\ \bibinfo {author} {\bibfnamefont {G.~G.}\ \bibnamefont
  {Naumis}},\ }\href@noop {} {\bibfield  {journal} {\bibinfo  {journal}
  {Physical Review B}\ }\textbf {\bibinfo {volume} {83}},\ \bibinfo {pages}
  {184204} (\bibinfo {year} {2011})}\BibitemShut {NoStop}%
\bibitem [{\citenamefont {McCann}, \citenamefont {Dykman},\ and\ \citenamefont
  {Golding}(1999)}]{mccann1999thermally}%
  \BibitemOpen
  \bibfield  {author} {\bibinfo {author} {\bibfnamefont {L.~I.}\ \bibnamefont
  {McCann}}, \bibinfo {author} {\bibfnamefont {M.}~\bibnamefont {Dykman}}, \
  and\ \bibinfo {author} {\bibfnamefont {B.}~\bibnamefont {Golding}},\
  }\href@noop {} {\bibfield  {journal} {\bibinfo  {journal} {Nature}\ }\textbf
  {\bibinfo {volume} {402}},\ \bibinfo {pages} {785} (\bibinfo {year}
  {1999})}\BibitemShut {NoStop}%
\bibitem [{\citenamefont {Simon}\ and\ \citenamefont
  {Libchaber}(1992)}]{simon1992escape}%
  \BibitemOpen
  \bibfield  {author} {\bibinfo {author} {\bibfnamefont {A.}~\bibnamefont
  {Simon}}\ and\ \bibinfo {author} {\bibfnamefont {A.}~\bibnamefont
  {Libchaber}},\ }\href@noop {} {\bibfield  {journal} {\bibinfo  {journal}
  {Physical review letters}\ }\textbf {\bibinfo {volume} {68}},\ \bibinfo
  {pages} {3375} (\bibinfo {year} {1992})}\BibitemShut {NoStop}%
\bibitem [{\citenamefont {Langer}, \citenamefont {Dorsey},\ and\ \citenamefont
  {Sethna}(1989)}]{langer1989entropy}%
  \BibitemOpen
  \bibfield  {author} {\bibinfo {author} {\bibfnamefont {S.~A.}\ \bibnamefont
  {Langer}}, \bibinfo {author} {\bibfnamefont {A.~T.}\ \bibnamefont {Dorsey}},
  \ and\ \bibinfo {author} {\bibfnamefont {J.~P.}\ \bibnamefont {Sethna}},\
  }\href@noop {} {\bibfield  {journal} {\bibinfo  {journal} {Physical Review
  B}\ }\textbf {\bibinfo {volume} {40}},\ \bibinfo {pages} {345} (\bibinfo
  {year} {1989})}\BibitemShut {NoStop}%
\bibitem [{\citenamefont {Naumis}(2012)}]{naumis}%
  \BibitemOpen
  \bibfield  {author} {\bibinfo {author} {\bibfnamefont {G.~G.}\ \bibnamefont
  {Naumis}},\ }\href@noop {} {\bibfield  {journal} {\bibinfo  {journal}
  {Physical Review E}\ }\textbf {\bibinfo {volume} {85}},\ \bibinfo {pages}
  {061505} (\bibinfo {year} {2012})}\BibitemShut {NoStop}%
\bibitem [{\citenamefont {Toledo-Mar{\'\i}n}, \citenamefont {Castillo},\ and\
  \citenamefont {Naumis}(2016)}]{toledo2016minimal}%
  \BibitemOpen
  \bibfield  {author} {\bibinfo {author} {\bibfnamefont {J.~Q.}\ \bibnamefont
  {Toledo-Mar{\'\i}n}}, \bibinfo {author} {\bibfnamefont {I.~P.}\ \bibnamefont
  {Castillo}}, \ and\ \bibinfo {author} {\bibfnamefont {G.~G.}\ \bibnamefont
  {Naumis}},\ }\href@noop {} {\bibfield  {journal} {\bibinfo  {journal}
  {Physica A: Statistical Mechanics and its Applications}\ }\textbf {\bibinfo
  {volume} {451}},\ \bibinfo {pages} {227} (\bibinfo {year}
  {2016})}\BibitemShut {NoStop}%
\bibitem [{\citenamefont {Wolynes}\ and\ \citenamefont
  {Lubchenko}(2012)}]{wolynes2012structural}%
  \BibitemOpen
  \bibfield  {author} {\bibinfo {author} {\bibfnamefont {P.~G.}\ \bibnamefont
  {Wolynes}}\ and\ \bibinfo {author} {\bibfnamefont {V.}~\bibnamefont
  {Lubchenko}},\ }\href@noop {} {\emph {\bibinfo {title} {Structural Glasses
  and Supercooled Liquids: Theory, Experiment, and Applications}}}\ (\bibinfo
  {publisher} {John Wiley \& Sons},\ \bibinfo {year} {2012})\BibitemShut
  {NoStop}%
\bibitem [{\citenamefont {Middleton}\ and\ \citenamefont
  {Wales}(2001)}]{Wales2}%
  \BibitemOpen
  \bibfield  {author} {\bibinfo {author} {\bibfnamefont {T.~F.}\ \bibnamefont
  {Middleton}}\ and\ \bibinfo {author} {\bibfnamefont {D.~J.}\ \bibnamefont
  {Wales}},\ }\href@noop {} {\bibfield  {journal} {\bibinfo  {journal}
  {Physical Review B}\ }\textbf {\bibinfo {volume} {64}},\ \bibinfo {pages}
  {024205} (\bibinfo {year} {2001})}\BibitemShut {NoStop}%
\bibitem [{\citenamefont {Wales}(2001)}]{wales2001microscopic}%
  \BibitemOpen
  \bibfield  {author} {\bibinfo {author} {\bibfnamefont {D.~J.}\ \bibnamefont
  {Wales}},\ }\href@noop {} {\bibfield  {journal} {\bibinfo  {journal}
  {Science}\ }\textbf {\bibinfo {volume} {293}},\ \bibinfo {pages} {2067}
  (\bibinfo {year} {2001})}\BibitemShut {NoStop}%
\bibitem [{\citenamefont {Hansen}\ \emph {et~al.}(2016)\citenamefont {Hansen},
  \citenamefont {Frick}, \citenamefont {Capaccioli}, \citenamefont {Hecksher},
  \citenamefont {Dyre},\ and\ \citenamefont {Niss}}]{hansen2016connection}%
  \BibitemOpen
  \bibfield  {author} {\bibinfo {author} {\bibfnamefont {H.~W.}\ \bibnamefont
  {Hansen}}, \bibinfo {author} {\bibfnamefont {B.}~\bibnamefont {Frick}},
  \bibinfo {author} {\bibfnamefont {S.}~\bibnamefont {Capaccioli}}, \bibinfo
  {author} {\bibfnamefont {T.}~\bibnamefont {Hecksher}}, \bibinfo {author}
  {\bibfnamefont {J.~C.}\ \bibnamefont {Dyre}}, \ and\ \bibinfo {author}
  {\bibfnamefont {K.}~\bibnamefont {Niss}},\ }\href@noop {} {\bibfield
  {journal} {\bibinfo  {journal} {arXiv preprint arXiv:1611.01748}\ } (\bibinfo
  {year} {2016})}\BibitemShut {NoStop}%
\bibitem [{Note1()}]{Note1}%
  \BibitemOpen
  \bibinfo {note} {Here and on, when taken the thermodynamic limit we will
  assume $\Omega _1/\Omega _0>1$.}\BibitemShut {Stop}%
\bibitem [{\citenamefont {Redner}(2001)}]{redner2001guide}%
  \BibitemOpen
  \bibfield  {author} {\bibinfo {author} {\bibfnamefont {S.}~\bibnamefont
  {Redner}},\ }\href@noop {} {\emph {\bibinfo {title} {A guide to first-passage
  processes}}}\ (\bibinfo  {publisher} {Cambridge University Press},\ \bibinfo
  {year} {2001})\BibitemShut {NoStop}%
\bibitem [{\citenamefont {Mel'nikov}\ and\ \citenamefont
  {Meshkov}(1986)}]{mel1986theory}%
  \BibitemOpen
  \bibfield  {author} {\bibinfo {author} {\bibfnamefont {V.}~\bibnamefont
  {Mel'nikov}}\ and\ \bibinfo {author} {\bibfnamefont {S.}~\bibnamefont
  {Meshkov}},\ }\href@noop {} {\bibfield  {journal} {\bibinfo  {journal} {The
  Journal of chemical physics}\ }\textbf {\bibinfo {volume} {85}},\ \bibinfo
  {pages} {1018} (\bibinfo {year} {1986})}\BibitemShut {NoStop}%
\bibitem [{\citenamefont {Gupta}\ and\ \citenamefont
  {Mauro}(2009)}]{Gupta2009}%
  \BibitemOpen
  \bibfield  {author} {\bibinfo {author} {\bibfnamefont {P.~K.}\ \bibnamefont
  {Gupta}}\ and\ \bibinfo {author} {\bibfnamefont {J.~C.}\ \bibnamefont
  {Mauro}},\ }\href@noop {} {\bibfield  {journal} {\bibinfo  {journal} {The
  Journal of chemical physics}\ }\textbf {\bibinfo {volume} {130}},\ \bibinfo
  {pages} {094503} (\bibinfo {year} {2009})}\BibitemShut {NoStop}%
\end{thebibliography}%

\end{document}